\documentclass[runningheads]{llncs}
\usepackage{graphicx}
\usepackage{amssymb, amsmath}
\begin{document}
	\title{TPRM: A Topic-based Personalized Ranking Model for Web Search}
	
	%
	%
	\author{Minghui Huang \and Wei Peng \and Dong Wang}
	%
	%
	\institute{Artificial Intelligence Application Research Center, Huawei Technologies, \\ 
		\email{\{huangminghui3, peng.wei1, wangdong153\}@huawei.com}\\
	}
	\maketitle              

	\begin{abstract}		
		Ranking models have achieved promising results, but it remains challenging to design personalized ranking systems to leverage user profiles and semantic representations between queries and documents. In this paper, we propose a {\bf t}opic-based {\bf p}ersonalized {\bf r}anking {\bf m}odel ({\bf TPRM}) that integrates user topical profile with pretrained contextualized term representations to tailor the general document ranking list. Experiments on the real-world dataset demonstrate that TPRM outperforms state-of-the-art ad-hoc ranking models and personalized ranking models significantly.
		
	\end{abstract}
	\section{Introduction}

	As the amount of information on the web grows rapidly, search engines are required to deal with large-scale natural language data. Ranking systems~\cite{DBLP:journals/ipm/GuoFPYAZWCC20}, which play a critical role in search engines, are required a deep understanding of the semantics behind queries and documents. Prior works~\cite{DBLP:conf/cikm/ShenHGDM14,DBLP:conf/cikm/GuoFAC16,DBLP:conf/sigir/XiongDCLP17} mainly focus on designing ranking systems by learning semantic matching between query and document terms. With the emergence of pretrained language models, i.e., BERT~\cite{DBLP:conf/naacl/DevlinCLT19}, existing ranking models benefit from contextual information learned from pretrained term representations~\cite{DBLP:journals/corr/abs-1904-07531,DBLP:conf/sigir/MacAvaneyYCG19}. Despite noticeable improvements in ranking performance, a BERT-based ranker focusing on pretrained surrounding contexts lacks effectiveness in modeling personalization signals, such as user clicks~\cite{DBLP:conf/www/DouSW07}. 
	
	On the other hand, personalized ranking takes a broader scope than user activities in human-computer interactions in searching and has been treated as user profiles~\cite{DBLP:conf/webi/SperettaG05,DBLP:conf/adaptive/GauchSCM07,DBLP:conf/wsdm/MatthijsR11}. Personalized ranking has been modeled in various ways, including traditional feature-engineered models~\cite{DBLP:conf/www/DouSW07,DBLP:conf/sigir/BennettWCDBBC12}, personalized topic models~\cite{DBLP:conf/cikm/HarveyCC13} and pure end-to-end neural architectures~\cite{DBLP:conf/cikm/GeDJNW18,DBLP:conf/sigir/LuDJNW19,DBLP:conf/cikm/MaDBW20,DBLP:conf/www/YaoDXW20} for ranking. Among these methods, we are interested in integrating the topic model approach to personalized ranking because a user profile can be succinctly captured by a set of easily explainable topics reflecting users' interests~\cite{DBLP:conf/www/MajumderS13}. Moreover, topic models help addressing the language discrepancy between queries and documents~\cite{DBLP:conf/sigir/GaoTY11,DBLP:conf/cikm/HuangHGDAH13,DBLP:conf/acl/PeineltNL20} by grouping different terms from a similar context into the same semantic cluster.

	
	~\author{DBLP:conf/sigir/ChiritaNPK05}~\cite{DBLP:conf/sigir/ChiritaNPK05} built a personalized ranking model based on the distance between a user profile defined using Open Directory Project (ODP) topics and the sets of ODP topics covered by each URL. Instead of employing a human-generated ontology, several latent topic models~\cite{DBLP:conf/cikm/CarmanCHB10,DBLP:conf/wsdm/HarveyRC11,DBLP:conf/cikm/HarveyCC13} based on Latent Dirichlet Allocation (LDA)~\cite{DBLP:conf/nips/BleiNJ01} are used to derive topic allocations for each of the documents to determine each user's topical interest profile. The produced ranking results are relevant to the user profile but with limited coverage of the semantics between queries and documents, resulting in sub-optimal performances. 
	
	In this paper, we propose a topic-based personalized ranking model ({\bf TPRM}) to utilize pretrained contextualized term representations and the user profiles constructed by a topic model. {\bf TPRM} leverages topic model-generated user profiles (based on clicked documents of each user) and pretrained contextual representations of the query and the candidate documents to tailor the output ranking list. The main contributions of this paper are:

	\begin{itemize}
		\item We integrate topic model-based user profile with pretrained language model to produce a novel personalized ranking system, outperforming state-of-the-art ad-hoc ranking models and personalized ranking models on a real-world AOL dataset~\cite{DBLP:conf/infoscale/PassCT06,DBLP:conf/sigir/AhmadCW19}. 
		\item We present the interpretability of user topical profiles by providing a means to visualize users preference in selecting documents under the given query.
		\item We disclose the effects of user interests and the semantic matching learned from queries and documents, revealing their positive contributions to the performance of TPRM.
	\end{itemize}
	
	
	\section{TPRM: The Proposed Model}
	\label{sec:model}
	\subsection{Problem Formulation}
	
	
	In a personalized ranking system,  documents are ranked based on both the user's interests and the given query. Specifically, let $q$ be a given query of user $u$ and let $d$ denote a document, the personalized ranking system aims at computing a relevance score $s(u, q, d)$ according to the user's interest, the query representation, and the document representation. And then, the candidate documents are ranked based on their relevance scores.
	
	\subsection{Architecture}
	
\begin{figure}
		\centering
		\includegraphics[width=\linewidth]{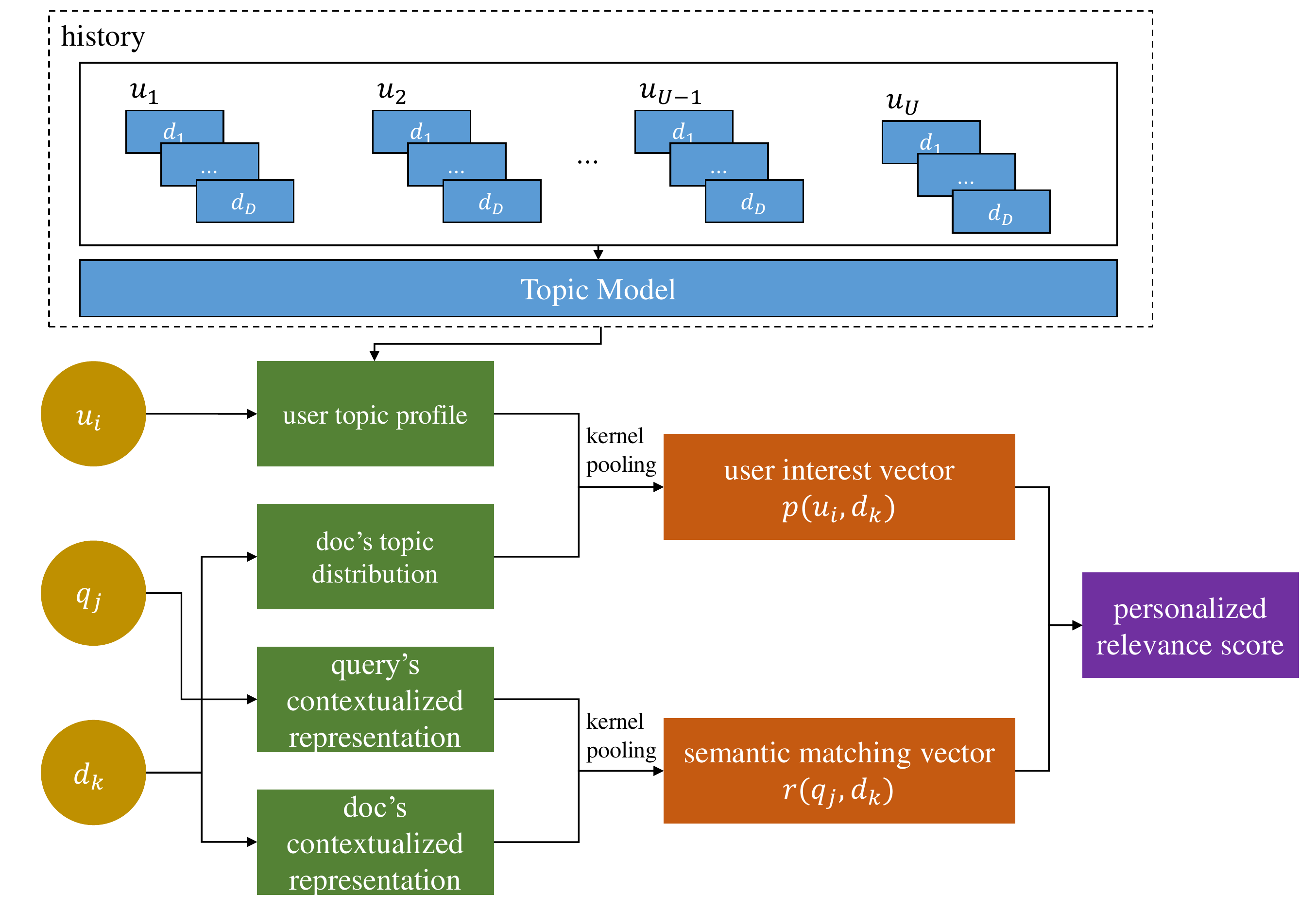}
		\caption{The architecture of the TPRM}
		\label{fig:TPRM}
	\end{figure}

	The architecture of our proposed TPRM model is shown in Figure~\ref{fig:TPRM}. The model mainly consists of four modules: (1) The user interest modeling, (2) The User-Doc interest matching, (3) The 		Query-Doc semantic matching, and (4) The personalized ranking.		
	\subsubsection{User Interest Modeling}
	User interests play a critical role in personalized ranking. To model user interest, we first build a topic model based on clicked documents in the search history. Then a document $d$ is represented by its topic distribution as follows,
	\begin{equation}
	\mathbf{t}_d = [t_{1,d}, t_{2,d}, \ldots, t_{T,d}] \in \mathbb{R}^{T},
	\end{equation} where $t_{i,d}$ is the $i$-th topic distribution of document $d$, and $T$ is the number of topics. Next, the user profile for user $u$ is constructed as:
	\begin{equation}
	\mathbf{p}_u = \mathrm{avg}\left(\{ \mathbf{t}_d, \, d \in \mathcal{D}_u \}\right) \in \mathbb{R}^{T},
	\end{equation} where $\mathcal{D}_{u}$ is the collection of clicked documents of user $u$. 
	
	\subsubsection{User-Doc Interest Matching}
	\label{sec:personal_interest}
	For a user $u$ and a candidate document $d$, we model the User-Doc interest matching via the kernel-pooling approach~\cite{DBLP:conf/sigir/XiongDCLP17}. Formally:
	\begin{eqnarray}
	M(u,d) &=& \cos(\mathbf{p}_u,\mathbf{t}_d), \\
	K_z(M(u, d)) &=& \exp(-(M(u,d)-\mu_z)^2/(2\sigma_z^2)), \\
	\mathbf{K} \big(M(u,d)\big) &=& [K_1(M(u,d)),\ldots, K_Z(M(u,d))], \\ 
	\theta(u,d) &=& \log(\mathbf{K}(M(u,d))),
	\end{eqnarray}
	where $M(u,d)$ is the translation of user profile and document topic, $K_z(M(u,d))$ denotes the $z$-th kernel, and $\theta(u,d) \in \mathbb{R}^Z$ is the user-doc interest matching vector.
	
	\subsubsection{Query-Doc Semantic Matching}
	Pretrained language models have been widely used in the field of information retrieval. We employ BERT to compute the semantic matching of a query and candidate documents. To be more specific, given a query $q$ with tokens $\{q_1, \ldots, q_m\}$ and a candidate document $d$ with tokens $\{d_1, \ldots, d_n\}$, we first concatenate $q$ and $d$ and then feed the sequence into BERT. Next, we derive the translation matrix $\mathbf{M}^l(q,d) \in \mathbb{R}^{m \times n}$ for the $l$-th layer of BERT as follows,
	\begin{equation}
	M^l_{ij}(q,d) = \cos(\mathbf{q}^l_i, \mathbf{d}^l_j),
	\end{equation}
	where $\mathbf{q}^l_i \in \mathbb{R}^{\mathrm{dim}}, \mathbf{d}^l_j \in \mathbb{R}^{\mathrm{dim}}$ are the contextualized representations for $q_i$ and $d_j$ in the $l$-th layer, respectively, $\mathrm{dim}$ is the dimension of BERT embedding, and $M^l_{ij}(q,d)$ is the $(ij)$-th element of matrix $\mathbf{M}^l(q,d)$. Next, similar to Section~\ref{sec:personal_interest}, we use the kernel-pooling technique with $Z$ kernels to derive the semantic matching vector as:
	\begin{eqnarray}
	K_z(M^l_i(q,d)) &=& \sum_{j}{\exp(-(M^l_{ij}(q,d)-\mu_z)^2/(2\sigma_z^2)}, \\
	\mathbf{K}(M^l_i(q,d)) &=& [K_1(M^l_i(q,d)),\ldots,K_Z(M^l_i(q,d))], \\
	\phi^l(q,d) &=& \sum_{i}{\log(\mathbf{K}(M^l_i(q,d))},  \\
	\phi(q,d) &=& [\phi^1(q,d);\ldots;\phi^L(q,d)],
	\end{eqnarray}
	where $;$ denotes the vector concatenation and $L$ is the number of BERT layers.	
	
	\subsubsection{Personalized Ranking}
	The personalized relevance score is computed with the user-doc and query-doc matching vectors as:
	\begin{equation}
	s(u,q,d) = \mathrm{MLP}([\theta(u,d);\phi(q,d)]),
	\end{equation}
	where $\mathrm{MLP}(\cdot)$ is a dense layer with the concatenated matching vectors as input and a scalar relevance score as output.

	\section{Experiment Settings}
	\label{sec:expeiment}
	\subsection{Datasets}
	We conduct experiments on the dataset constructed by~\author{DBLP:conf/sigir/AhmadCW19}~\cite{DBLP:conf/sigir/AhmadCW19} using the real-world AOL search log~\cite{DBLP:conf/infoscale/PassCT06}. Following previous works~\cite{DBLP:conf/sigir/AhmadCW19}, search logs in the first five weeks are set as history, and the remaining data are used for model training, validation and testing with a ratio of 6:1:1.

	
	
	\subsection{Evaluation Metrics}
	We use Mean Average Precision (MAP), Mean Reciprocal Rank (MRR), P@1 (precision in the first positions) and A.Clk (average click position)~\cite{DBLP:conf/cikm/GeDJNW18,DBLP:conf/sigir/LuDXWW20} metrics to evaluate the quality of the ranking list.
	
	\subsection{Settings}
	\subsubsection{Topic Model}
	Our experiments are implemented on the Latent Dirichlet Allocation (LDA)~\cite{DBLP:conf/nips/BleiNJ01} approach since it is the most popular and widely used topic model. The number of topics is an important hyper-parameter in topic models. To identify a useful topic model for our TPRM, we build LDA with different numbers of topics and evaluated them with multiple metrics. Our experiments in Section~\ref{sec:topicnumber} show that LDA with 50 topics has the best performance.
	
	\subsubsection{Kernel Pooling Layer}
	Following~\author{DBLP:conf/sigir/XiongDCLP17}~\cite{DBLP:conf/sigir/XiongDCLP17}, we use 11 kernels in the kernel pooling layer. $\mu \in \{-0.9,-0.7,...,0.9\}$ and $\sigma=0.1$ are set for the first 10 kernels, while $\mu_0 = 1.0$ and $\sigma = 1e-3$ are set for the last kernel to harvest exact matches.
	
	\subsubsection{Training}
	We use the BERT base model in our experiments and set maximum allowable length of query and document to be 10 and 500 respectively. We train the model using the Adam optimizer and pairwise hinge loss, with a learning rate $1e-3$ for the model and $2e-5$ for BERT layers~\cite{DBLP:conf/sigir/MacAvaneyYCG19}. We train the model for 10 epochs, each with 16 batches of 8 training pairs, and employ MRR to select the best performing model on the validation set.

	\subsection{Baselines}
	{\bf TPRM} uses the pretrained contextualized term representations of BERT to learn semantic matching between queries and documents. We evaluate the performance of our model by comparing it with {\bf BM25} algorithm~\cite{DBLP:conf/trec/RobertsonWJHG94} and several state-of-the-art ad-hoc ranking models using semantic matching.
	
	\begin{itemize} 
		\item {\bf KNRM}~\cite{DBLP:conf/sigir/XiongDCLP17} is an ad-hoc interactive matching model, which uses kernel-pooling upon the word-level embedding translation matrix to calculate query-document relevance.
		\item {\bf Conv-KNRM}~\cite{DBLP:conf/wsdm/DaiXC018} is enhanced from the KNRM model, which adds a convolutional layer for modeling n-gram soft matches. It integrates local contextual information to improve the ranking performance.
		\item {\bf CEDR-KNRM}~\cite{DBLP:conf/sigir/MacAvaneyYCG19} uses KNRM to extract features from each layer of BERT's token vectors, and then these features are incorporated with BERT's classification vector to generate the final ranking list.

	\end{itemize}
	
	Our model also uses topic models to construct user profiles for ranking personalization, hence we employ several state-of-the-art personalized ranking models as baselines. HRNN, PSGAN and PSTIE are based on recurrent neural network, which is hard to model long sequences, hence they only use the title filed.
	
	\begin{itemize} 
		\item {\bf P-Click}~\cite{DBLP:conf/www/DouSW07} generates the final personalized ranking list by counting the number of the user's clicks under the same queries.		
		\item {\bf SLTB}~\cite{DBLP:conf/sigir/BennettWCDBBC12} learns to combine 102 long-term and short-term features for each query, and then uses the LambdaMart~\cite{wu2008ranking} to generate the final personalized ranking list.
		\item {\bf HRNN}~\cite{DBLP:conf/cikm/GeDJNW18} uses a hierarchical recurrent neural network with a query-aware attention model to build user profiles dynamically for personalized ranking.		
		\item {\bf PSGAN}~\cite{DBLP:conf/sigir/LuDJNW19} is a personalized framework that applies a generative adversarial network (GAN) to return a better personalized ranking list. In this paper, we choose the document-selection based model as the baseline.
		\item {\bf PSTIE}~\cite{DBLP:conf/cikm/MaDBW20} is a fine-grained time information enhanced model for personalized ranking, which constructs user interest representations using the time information associated with user actions. These representations are then used as the input of the MV-LSTM to calculate the matching score between query and candidate document.
	\end{itemize}

	\section{Result and Analysis}
	\label{sec:result}
	\subsection{Results}
	\begin{table}
		\caption{Results of ad-hoc ranking models on the AOL dataset.}
		\label{tab:result_adhoc}
		\centering
		\begin{tabular}{|c|c|c|c|c|}
			\hline
			&MAP&MRR&P@1&A.Clk\\
			\hline
			BM25&0.250 &0.258 &0.148&17.152 \\
			\hline
			KNRM&0.429 &0.439&0.270 &9.941 \\
			\hline
			Conv-KNRM&0.474 &0.485 &0.327 &9.471\\
			\hline
			CEDR-KNRM&0.546&0.544&0.440&8.751\\
			\hline
			TPRM-semantic&0.543&0.542&0.436&8.956\\			
			\hline
		\end{tabular}
	\end{table}

	\begin{table}
	\caption{Results of personalized ranking models on the AOL dataset.}
	\label{tab:result}
	\centering
	\begin{tabular}{|c|c|c|c|c|}
		\hline
		&MAP&MRR&P@1&A.Clk\\
		\hline
		P-Click&0.422 &0.430 &0.379 &16.526 \\
		\hline
		SLTB&0.507 &0.519&0.466 &13.926 \\
		\hline
		HRNN&0.542 &0.555&0.485 &10.552 \\
		\hline
		PSGAN&0.548 &0.560 &0.489&10.267 \\
		\hline
		PSTIE&0.564&0.577& 0.503&-\\
		\hline
		TPRM&{\bf 0.599}&{\bf0.597}&{\bf0.505}&{\bf7.50}\\
		
		\hline
	\end{tabular}
\end{table}

	We compare our model with the baselines on the AOL dataset, and present results in Table~\ref{tab:result_adhoc} and Table~\ref{tab:result}. Results demonstrate that our model can significantly improve ad-hoc ranking models and personalized ranking models. Meanwhile, {CEDR-KNRM} greatly outperforms other ad-hoc models, verifying that pretrained contextualized term representations can significantly contribute to ranking systems. Moreover, most personalized ranking models outperform ad-hoc ranking models, indicating the effectiveness of user profiles for ranking systems. We also show the performance of {\bf TPRM-semantic}, which is the TPRM without user interest component. TPRM significantly outperforms TPRM-semantic, verifying the benefits of user profiles constructed by the topic model.
	
	\subsection{Analysis of User Topical Profile}
	\subsubsection{Topic Number}
	\label{sec:topicnumber}
	\begin{figure}
		\centering
		\includegraphics[width=\linewidth]{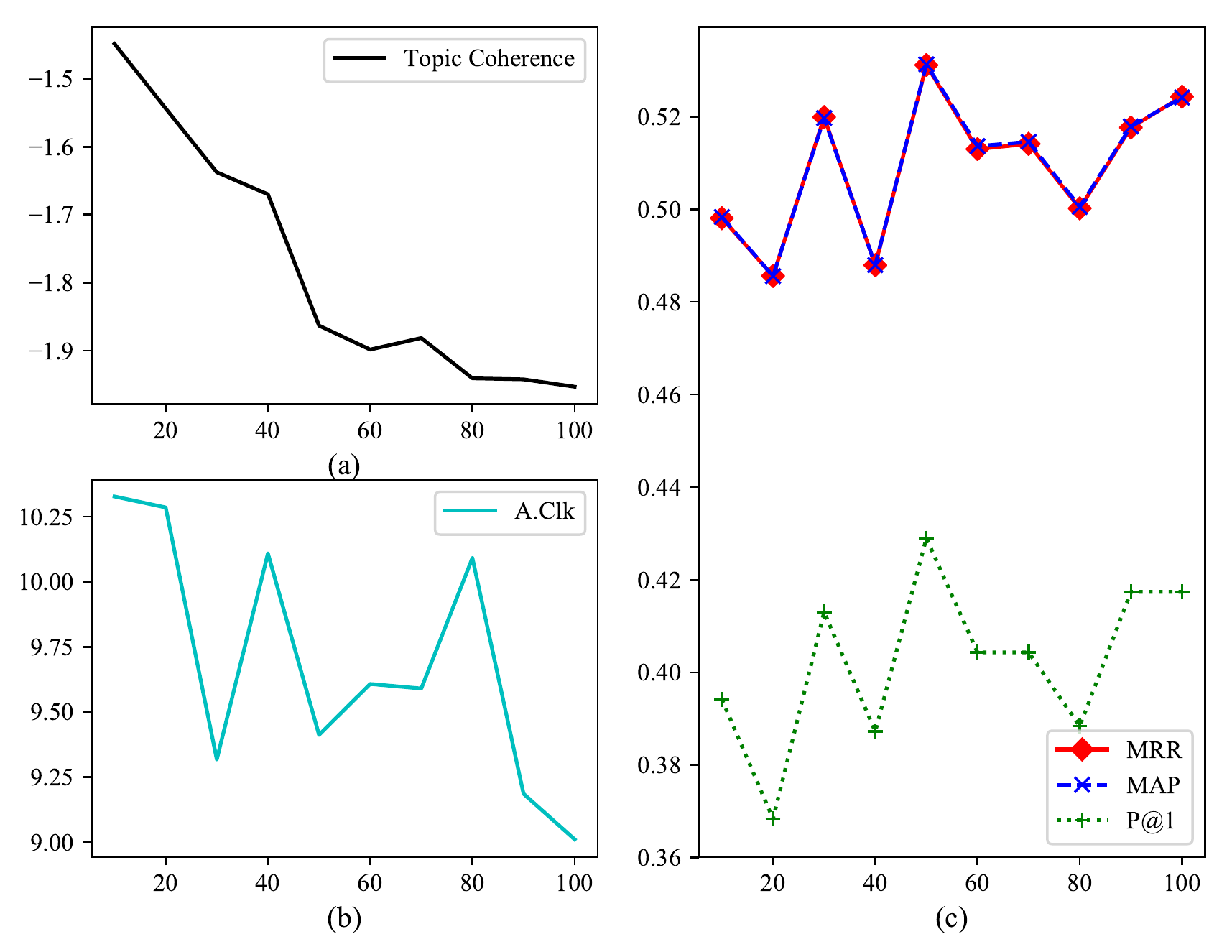}
		\caption{(a) Topic coherence of the topic model with different numbers of discovered topics. Results of the TPRM with different topic number on the sampled AOL dataset: (b) A.Clk; (c) MRR, MAP and P@1.}
		\label{fig:topicnumber}
	\end{figure}
	The number of topics is an important hyper-parameter in topic models. To investigate the quality of topics discovered by the topic model, we use the topic coherence score~\cite{DBLP:conf/wsdm/RoderBH15} as the evaluation metric. Intuitively, a higher topic coherence score indicates better quality of topics. Results are shown in the Figure~\ref{fig:topicnumber}(a), which demonstrates that the quality of discovered topics improves rapidly with increasing the number of topics until 50 when it is stabilized. Moreover, to identify a useful topic model for the dataset, we sample 1\% data from the train, validation and test sets separately to obtain an effective topic number without expensive training on the full dataset. Results of our experiments are shown in Figure~\ref{fig:topicnumber}(b) and Figure~\ref{fig:topicnumber}(c), which indicates that the TPRM has the best performance when the number of topics is 50.
	
	\subsubsection{Interpretability}
	\begin{figure}
		\centering
		\includegraphics[width=0.8\linewidth]{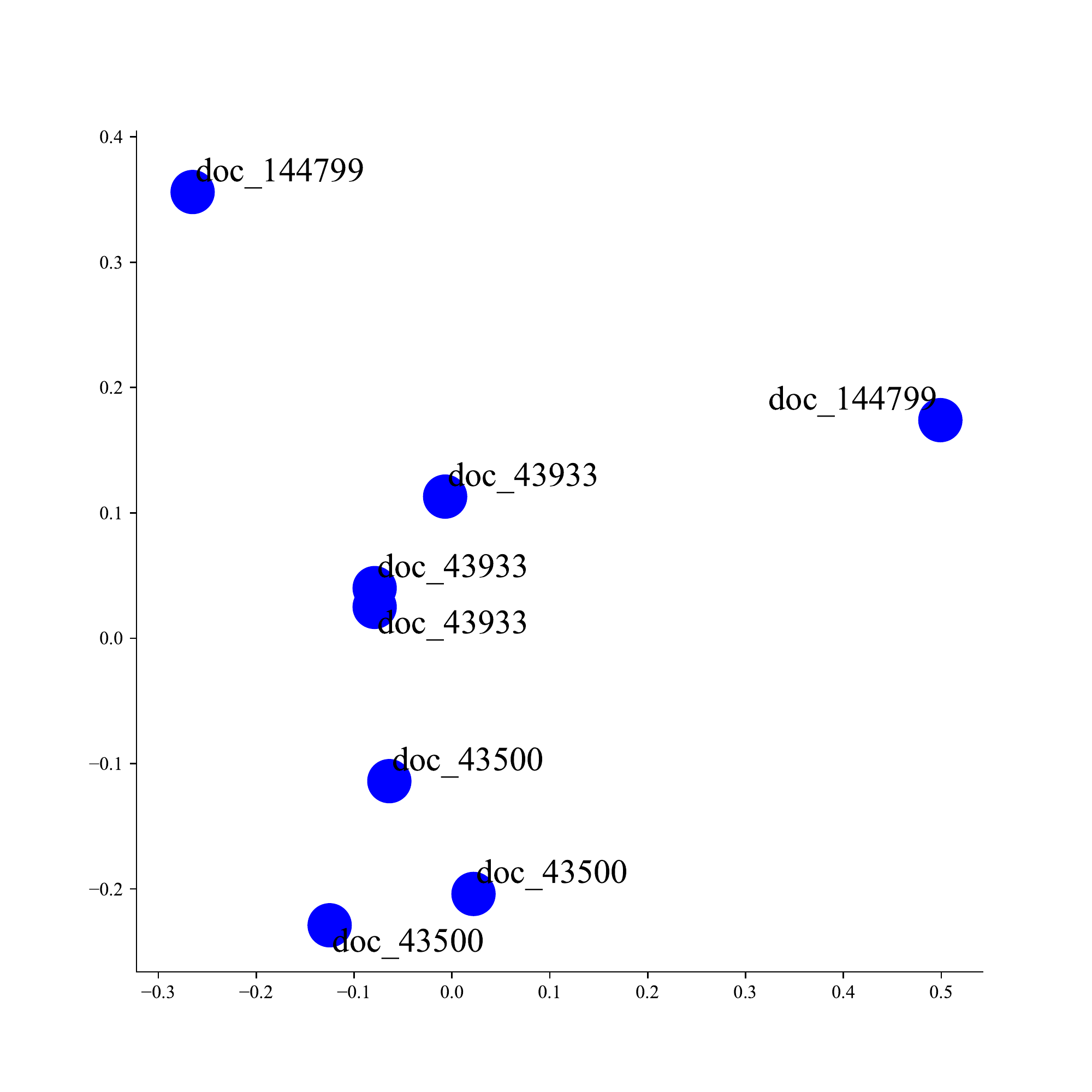}
		\caption{Scatter plot of user topical profiles and the user's clicked documents, where each point indicates a user and the text annotation of each point indicates the document clicked by the user. }
		\label{fig:scatter}
	\end{figure}
	
	To further evaluate the interpretability of user profiles, we map $\mathrm{profile}(u)$ onto a two-dimensional space via Principal Component Analysis (PCA).	Figure~\ref{fig:scatter} presents an example of 8 users, who have given the query ``suzuki'' and clicked 3 different documents (``doc\_144799'', ``doc\_43933'' and ``doc\_43500'') in the test set. In the figure, each point denotes a user, and the text annotation is the clicked documents for different users under the same query. The scatter plot indicates that users having the same clicked document are closer than those of different clicked documents.
	
	\subsection{Analysis of User Interest}
	\begin{figure}
		\centering
		\includegraphics[width=0.9\linewidth]{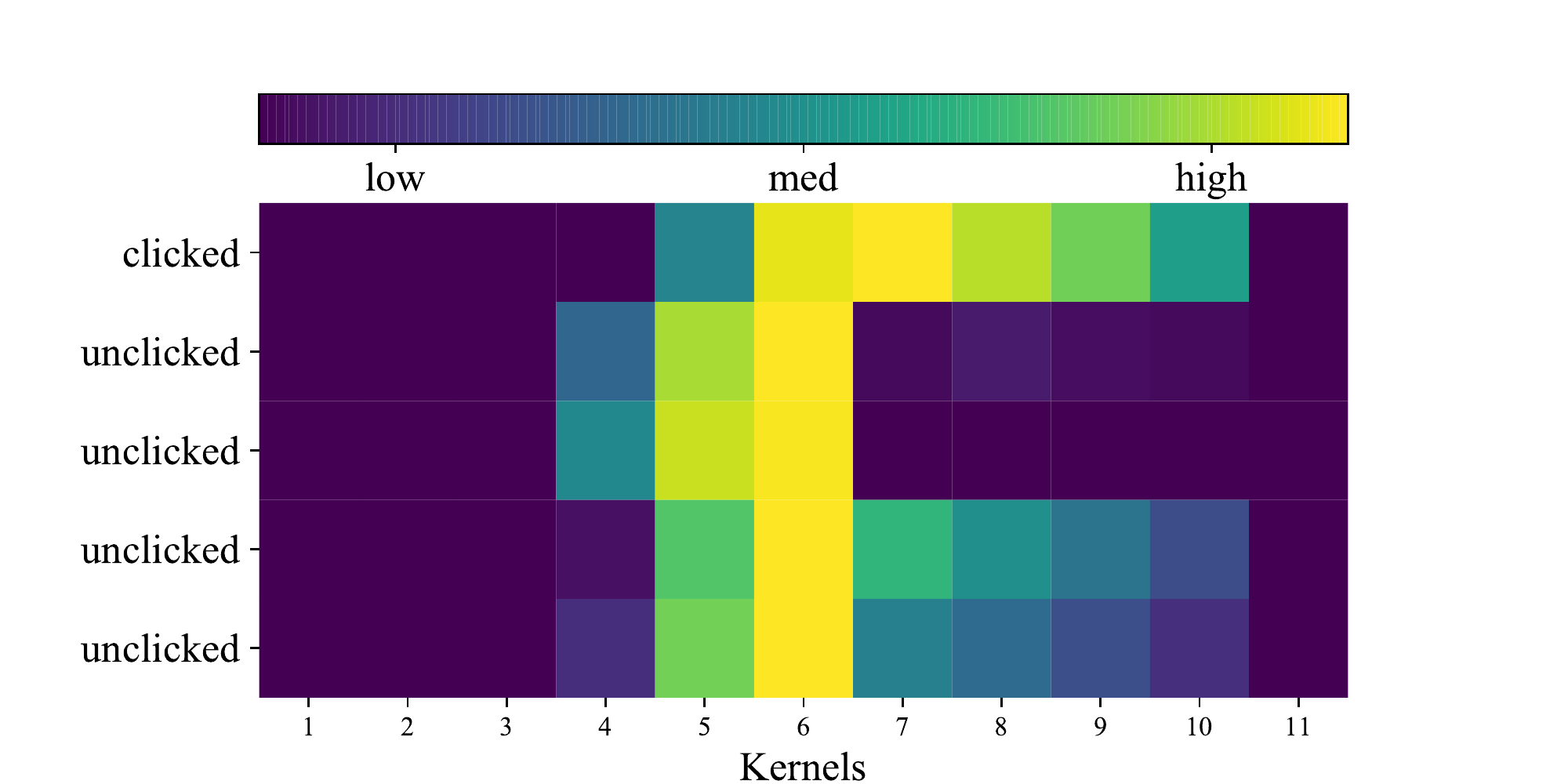}
		\caption{Heatmap of user interests for the query ``propaganda bandwagon stereotype''.}
		\label{fig:heatmap}
	\end{figure}
	In this part, we conduct a qualitative inspection of user interests used in the TPRM. Figure~\ref{fig:heatmap} shows values of 11 kernels in the user interest component (Section~\ref{sec:personal_interest}) with the query ``propaganda bandwagon stereotype''. To better display, we sample 1 clicked document and 4 unclicked documents of the query. Centers of the first 10 kernels are from $-0.9$ to $0.9$. From this heatmap, we can find that user interest scores are higher in clicked documents, which indicates that kernels in the user interest component are more activated in clicked documents. It demonstrates that the user interest component is helpful to distinguish relevant from irrelevant documents.	
	
	\subsection{Analysis of Semantic Matches}
	\begin{figure}
		\centering
		\includegraphics[width=0.7\linewidth]{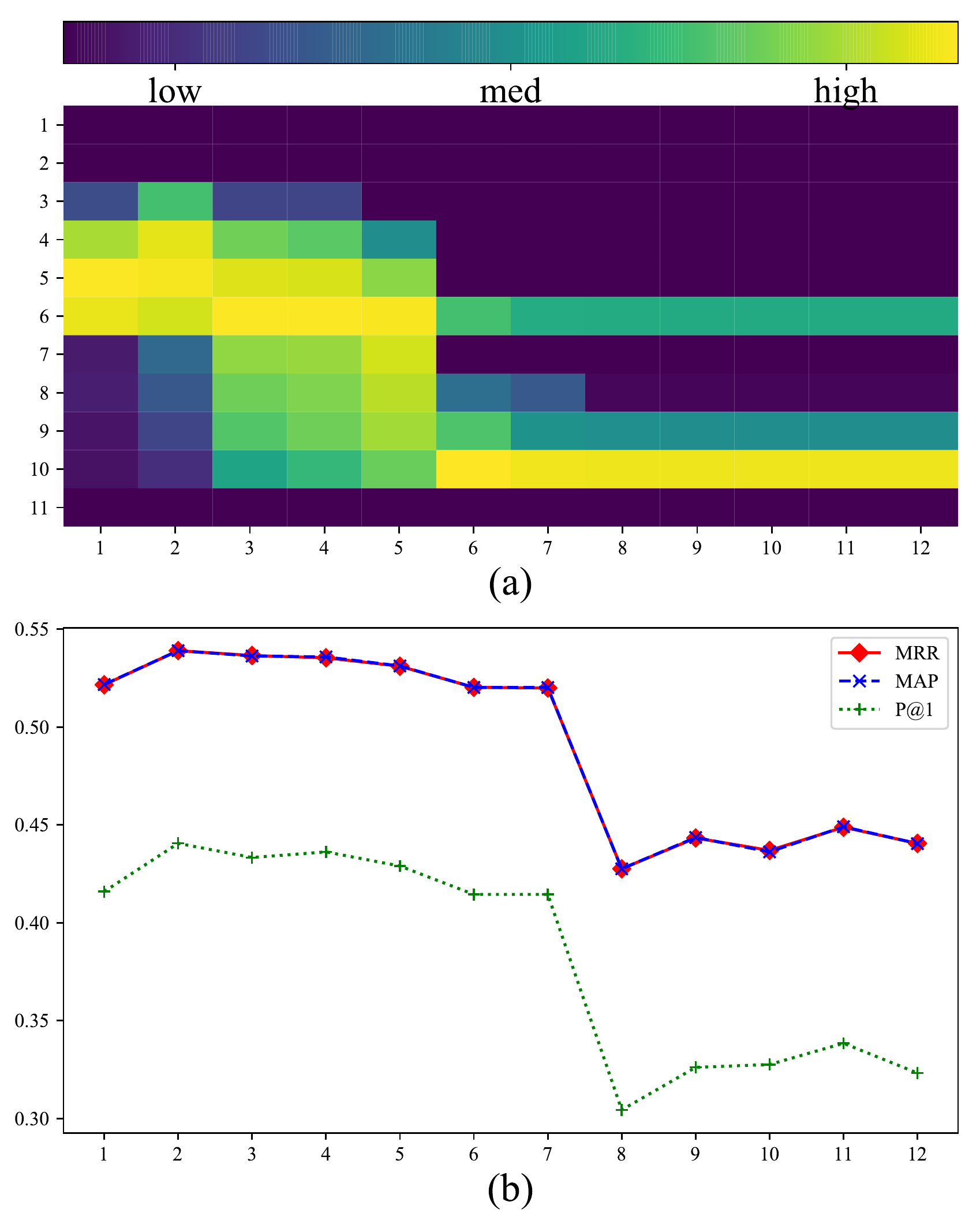}
		\caption{(a) Values of 11 kernels in different BERT layers. (b) Results of the TPRM with different BERT layers.}
		\label{fig:layers}
	\end{figure}
	To further analyze the effects of different BERT layers, we draw a heatmap in Figure~\ref{fig:layers}(a) for the query ``propaganda bandwagon stereotype'', where X-axis marks layer levels from shallow (1) to deep (12) and Y-axis marks 11 kernels in the semantic matching component. From the figure, we can find that shallow BERT layers have higher values in kernels, which means that these layers are more activated. To investigate the effects of different BERT layers, we conduct experiments on the sampled dataset with different BERT layers. Results of the TPRM with different BERT layers are shown in Figure~\ref{fig:layers}(b). The performances of the TPRM with BERT layers are better in the shallow layers and drop rapidly in the last few layers, which is also indicated in~\cite{DBLP:conf/sigir/ZhanMLZM20} that shallow BERT layers have better ranking performance than the last few layers. In shallow BERT layers, the semantic matching component is more activated, the performance of the TPRM is better, which verified that the semantic matching component is highly contributed to the TPRM.
	\section{Conclusion}
	\label{sec:conculusion}
	We propose a personalized ranking model integrating topical user profile with a pretrained contextual language model in this work. Experiments demonstrate that our model can significantly outperform ad-hoc ranking models and personalized ranking models. The study show users with similar topical profiles tend to select the same document under the same query. Ablation studies and our further analysis reveal the effects of user interests and the semantic matching learned from queries and documents. 
	
	%
	%
	\bibliographystyle{splncs04}
	\bibliography{samplepaper}
	
\end{document}